\documentclass[prd,reprint,aps,amsmath,amssymb,showpacs,nofootinbib]{revtex4-1}
\usepackage[utf8x]{inputenc}
\usepackage{graphicx}
\usepackage{enumitem}
\usepackage{slashed}
\usepackage{url}
\usepackage[pdftex]{hyperref}
\usepackage[T1]{fontenc}
\usepackage{bm}
\usepackage{color}

\begin{document}

\title{Probing nonstandard neutrino cosmology with terrestrial neutrino experiments}
\author{Akshay Ghalsasi}
\email{aghalsa2@uw.edu}
\affiliation{Department of Physics, University of Washington, Seattle, WA 98195, USA}

\author{David McKeen}
\email{dmckeen@uw.edu}
\affiliation{Department of Physics, University of Washington, Seattle, WA 98195, USA}

\author{Ann E. Nelson}
\email{aenelson@uw.edu}
\affiliation{Department of Physics, University of Washington, Seattle, WA 98195, USA}                             
\date{\today}

\begin{abstract}
Neutrino masses and the number of light neutrino species can be tested in a variety of laboratory experiments and also can be constrained by particle astrophysics and precision cosmology. A conflict between these various results could be an indication of new physics in the neutrino sector. In this paper we explore the possibility for reconciliation of otherwise discrepant results in a simple model containing a light scalar field which produces Mass Varying Neutrinos (MaVaNs). We extend previous work on MaVaNs to consider issues of neutrino clumping, the effects of additional contributions to neutrino mass, and   reconciliation of eV mass sterile neutrinos with cosmology.
\end{abstract}

\maketitle

\section{Introduction}
\label{sec:intro}
Over the past twenty years, definitive evidence for neutrino oscillations from a host of experiments has revealed that neutrino masses are nonzero. Because neutrino masses cannot be accounted for in the standard model (SM), this is a clue to physics beyond the SM. Since oscillation experiments are only sensitive to the difference in (squared) masses, the overall mass scale is not known and only two mass differences have been conclusively established, the so-called solar  and atmospheric mass splittings, $\Delta m_{\odot}^2\simeq7.5\times10^{-5}~\rm eV^2$ and $\Delta m_{\rm atm}^2\simeq2.4\times10^{-3}~\rm eV^2$. It has also been well established that the neutrino mixing matrix, characterizing the mismatch between weak interaction and mass eigenstates, involves large mixing angles, in stark contrast to the quark sector. While they may seem like an uninteresting example of new physics---we have seen other chiral fermions obtain masses in the SM---neutrino masses could differ in a fundamental way from other fermion masses. Because neutrinos are not charged under electromagnetism, their mass generation mechanism could involve Majorana masses, violating lepton number, while the nonzero charges of the other fermions requires their masses to be of purely Dirac form. To generate neutrino masses in a way that does not disturb the successful picture we have of electroweak symmetry breaking, new neutrino states that are uncharged under the electroweak gauge group (or ``sterile'' neutrinos, as opposed to the ``active'' ones that carry electroweak charge) are typically invoked. Since they are gauge singlets, mass terms for these sterile neutrinos need not involve Higgs fields, which means that the mass scale in the sterile neutrino sector is largely a free parameter. While there may be theoretical bias for this scale to be very large compared to the weak scale, it is entirely possible and self-consistent that it is within reach of current experiments.

Indeed, there are phenomenological reasons to consider a mass scale in the sterile neutrino sector as small as an eV. Alongside this standard three neutrino picture, there have been a number of experimental hints of neutrino oscillations characterized by a squared mass splitting of $\Delta m^2\sim{\cal O}\left(1~\rm eV^2\right)$ and a mixing angle $\theta\sim{\cal O}\left(0.1\right)$; these include short-baseline reactor experiments~\cite{Mueller:2011nm,*Huber:2011wv}, the flux of neutrinos from radioactive sources in gallium solar neutrino experiments~\cite{Acero:2007su,*Giunti:2010zu}, and electron (anti)neutrino appearance in muon (anti)neutrino beams~\cite{Aguilar:2001ty,*Aguilar-Arevalo:2013pmq}. To interpret these data in terms of neutrino oscillations requires an additional (sterile) neutrino around an eV and a large mixing angle with the active neutrinos. For detailed analyses, see, e.g.,~\cite{Kopp:2013vaa,*Gonzalez-Garcia:2015qrr}. It should be noted that there is generally tension between disappearance and appearance data, as a recent search for $\nu_\mu$ and $\bar\nu_\mu$ disappearance at IceCube~\cite{TheIceCube:2016oqi} shows, disfavoring the sterile neutrino interpretation of electron (anti)neutrino appearance data. However, a global fit including the IceCube results claims that a relatively large active-sterile mixing is allowed~\cite{Collin:2016aqd}.

Additionally, progress in the direct search for neutrino masses has been ongoing. Searches using the endpoint in tritium $\beta$-decay currently limit the electron neutrino mass to less $2.05~\rm eV$ at 95\% CL~\cite{Aseev:2011dq}. The upcoming KATRIN~\cite{Robertson:2013ziv} and Project-8~\cite{Doe:2013jfe} experiments hope to probe masses down to about $0.1-0.2~\rm eV$ in the near future. Conceivably, the PTOLEMY experiment could use inverse beta decay to be sensitive to cosmological neutrinos~\cite{Weinberg:1962zza,Betts:2013uya,Long:2014zva}.

In parallel with  progress in neutrino measurements, cosmology has entered an era of impressive precision, enabling cosmological tests of physics beyond the standard model. We have direct observational evidence of the state of the Universe up to temperatures of a few MeV, corresponding to the time of neutrino decoupling and primordial nucleosynthesis (BBN). The precise picture of the Universe we now have at these temperatures and below allows for new physics below an MeV, even if weakly coupled, to be confronted with observation.

Because (active or sterile) neutrinos interact very weakly with the rest of the Universe after decoupling--acting as a form of noninteracting radiation until they become nonrelativistic when they begin to act like dark matter--their observational consequences are relatively easy to understand. At early times, the cosmic microwave background (CMB), structure formation, and BBN are all sensitive to the energy density in neutrinos, which can be related to their masses and, in the case of sterile neutrinos, their mixing with the active neutrinos. The general agreement of these data with the standard cosmological picture based on three (essentially massless) neutrinos allows constraints to be placed on additional sterile neutrinos or on the masses of the active neutrinos. However, at late times, the neutrino energy density is the only SM component that can have nonstandard cosmology, which makes finding probes of this behavior crucial.

In the case of a single massive sterile neutrino that is fully thermalized at early times, an up-to-date fit to cosmological observations give an upper bound on its mass (assuming the light, mostly active neutrinos' masses are negligible) of $0.53~\rm eV$~\cite{Archidiacono:2016kkh}. In the standard case of only three (active) neutrinos, the Planck analysis of only CMB data constrains the sum of the light neutrino masses to $0.675~\rm eV$~\cite{Ade:2015xua}. Including further cosmological data improves the bounds to $0.3~\rm eV$~\cite{Ade:2015xua,Thomas:2009ae,*RiemerSorensen:2011fe,*Zhao:2012xw}. Note that these upper limits are all at 95\%~CL. Improved observations could allow values of the sum of the active neutrino masses as small as $0.06~\rm eV$ to be probed~\cite{Feng:2014uja}. Sterile neutrinos which are much lighter than an eV and have similar abundance to the active neutrinos are disfavored by the Planck determination of $N_{\rm eff}$~\cite{Ade:2015xua}.

At first glance, the null results from cosmological analyses are in strong tension with the sterile neutrino interpretation of the short baseline anomalies. In addition, the projected reach in the limit on the sum of the neutrino masses in the standard three neutrino scenario coming from cosmology seems to imply that current laboratory searches will not be sensitive enough to see nonzero neutrino masses. However, these conclusions rely on the assumption of a standard cosmological history. Thus, one should view contemporary terrestrial experiments seeking to measure neutrino masses or to test the sterile neutrino solution to short baseline anomalies as nontrivial probes of cosmology. Some scenarios that allow for cosmological observations to be compatible with eV mass neutrinos include coupling the sterile (with respect to the SM) neutrinos to a new U(1) gauge boson~\cite{Hannestad:2013ana,*Dasgupta:2013zpn,*Bringmann:2013vra,*Saviano:2014esa,*Chu:2015ipa,*Cherry:2016jol} or pseudoscalar~\cite{Archidiacono:2014nda,*Archidiacono:2016kkh}, or allowing the sterile neutrinos to be chiral under a new gauge group~\cite{Berezhiani:1995yi,*Vecchi:2016lty}. 

In this paper we will  focus on the reconciliation of   eV mass neutrinos with cosmology via the dependence of  the neutrino masses and mixing angles  on the expectation value of a non constant light scalar field.  This possibility  was originally motivated to address the puzzle of dark energy~\cite{Fardon:2003eh,Fardon:2005wc}, but we will consider this possibility more generally, including the possibility of additional contributions to neutrino mass, the effects of  neutrino clustering, and models which do not give dark energy. We will consider two scenarios. In~\S~\ref{sec:logpotential} we consider a MaVaN model containing a light scalar field with a logarithmic potential, and find parameters such that eV mass sterile neutrinos with sizable mixing angles are allowed today which were always heavy enough at earlier times so that these states were never populated in the early universe and have no observable effect on cosmology. In~\S~\ref{sec:mass} we consider a scenario which allows the observed, active neutrinos to have a mass which is today around an eV. Because the masses were much lighter at high redshift, cosmological observations indicate a much smaller mass. In~\S~\ref{sec:susy} we discuss a cosmologically viable supersymmetric MaVaN scenario which could allow eV mass sterile neutrinos to appear in terrestrial experiments.  

\section{A logarithmic potential and an $\rm eV$ sterile neutrino}
\label{sec:logpotential}

\subsection{A single active neutrino}
 
We begin by describing a framework with one active flavor,  and will  discuss incorporating three flavors in~\S~\ref{sec:3nus}. We introduce an active (i.e. electroweak doublet) neutrino, $\nu$, and a sterile (i.e. electroweak singlet) neutrino, $N$. After electroweak symmetry breaking, their masses are generated by
\begin{align}
{\cal L}_{\rm mass}&=-m_D\nu N-m_N N N+{\rm h.c.}
\label{eq:simpleL}
\end{align}
As is well known, in the limit $m_D\ll m_N$, this leads to a light, mostly active neutrino, $\hat \nu=\nu+\theta N$, with a mass $m\simeq m_D^2/m_N$ and a heavy, mostly sterile neutrino, $\hat N=N-\theta\nu$, with mass $M\simeq m_N$. (We use hats here and below to denote mass eigenstates.) The active-sterile mixing angle is $\theta\simeq\sqrt{m/M}$.

The short baseline reactor anomaly suggests oscillations between active (electron in this case) and sterile neutrinos with a mixing angle of ${\cal O}\left(0.1\right)$ and a squared mass splitting of ${\cal O}\left(1~\rm eV^2\right)$. This can easily be accounted for by choosing $m\sim 0.01~\rm eV$ and $M\sim 1~\rm eV$. This simple explanation of short baseline anomalies is in tension with cosmological observations because the heavy neutrino with a mass $\sim 1~\rm eV$ will be in thermal equilibrium at the time neutrinos decouple, due to its relatively large admixture of active neutrino~\cite{Enqvist:1991qj,*Barbieri:1989ti,*Kainulainen:1990ds,*DiBari:2001ua,*Abazajian:2002bj,*Dolgov:2002wy,*Dolgov:2003sg,*Cirelli:2004cz,*Dodelson:2005tp,*Dolgov:2008hz}.

However, as mentioned above there are well-motivated scenarios where this conclusion does not hold. One of the simplest possibilities is when the sterile neutrino's Majorana mass depends on  the value of a scalar field which we call $A$. Because the light neutrino mass is determined by the Majorana mass, it also depends on $A$, and therefore a finite density background of light neutrinos can give corrections to the potential for $A$. Since the density of neutrinos is determined by the temperature, these corrections can cause the value of $A$ to vary with temperature (or, equivalently, time, as the Universe cools).

To see this, consider the contribution at a temperature $T$ to the energy density from the light neutrino, whose mass $m\left(A\right)$ varies with the scalar field $A$,
\begin{align}
\delta V\left(A,T\right)=2\times\int \frac{d^3p}{\left(2\pi\right)^3}\frac{\sqrt{p^2+m^2\left(A\right)}}{e^{p/T}+1}.
\label{eq:deltaV}
\end{align}
The effective potential for $A$ is given by its zero temperature scalar potential, $V_0$, and the contribution from the neutrino background,
\begin{align}
V\left(A,T\right)&=V_0\left(A\right)+\delta V\left(A,T\right).
\label{eq:Vfull}
\end{align}
We first consider a logarithmic scalar potential~\cite{Fardon:2003eh},
\begin{align}
V_0&=\Lambda^4\log\left(1+\left|\frac{A}{\sigma}\right|\right),
\label{eq:Vlog}
\end{align}
with $\sigma$ small compared to the range of relevant $A$ values. We will discuss a quadratic potential in~\S~\ref{sec:susy} when we introduce the SUSY version of theory. Taking the light neutrino to be relativistic, the full scalar potential at finite $T$ is
\begin{align}
V\left(A,T\right)&=\Lambda^4\log\left(1+\left|\frac{A}{\sigma}\right|\right)+\frac{m^2\left(A\right)T^2}{24}+{\cal O}\left(m^4\left(A\right)\right).
\label{eq:Vfull1}
\end{align}
We assume, quite generally, that the sterile neutrino mass has $A$-dependent and -independent terms,
\begin{align}
m_N\left(A\right)&=m_0+\kappa A.
\label{eq:Mmajorana}
\end{align}
In this case, the light neutrino mass is $m_{\nu}\left(A\right)=m_D^2/\left(m_0+\kappa A\right)$, resulting in an effective potential of
\begin{align}
V\left(A,T\right)&=\Lambda^4\log\left(1+\left|\frac{A}{\sigma}\right|\right)+\frac{m_D^4T^2}{24\left(m_0+\kappa A\right)^2}.
\label{eq:Vfull2}
\end{align}
The first term tends to push the scalar field toward smaller values, while the second (the importance of which increases at high $T$) prefers larger values of $A$; the interplay of the two determines the value of $A$ that minimizes the effective potential. At high temperatures, $\kappa A$ can be large compared to $m_0$. When this is the case, minimizing the effective potential results in $A\propto T$ so that the heavy (mostly sterile) neutrino tracks the temperature, $M\propto T$. The light neutrinos mass is therefore smaller at large temperatures, $m_{\nu}\propto T^{-1}$. At some temperature, $\kappa A$ becomes comparable to $m_0$. $A$ then moves toward its minimum as determined by $V_0$ and the neutrino masses approach the temperature-independent values $M\simeq m_0$ and $m\simeq m_D^2/m_0$.

\begin{figure}
\includegraphics[width=\linewidth]{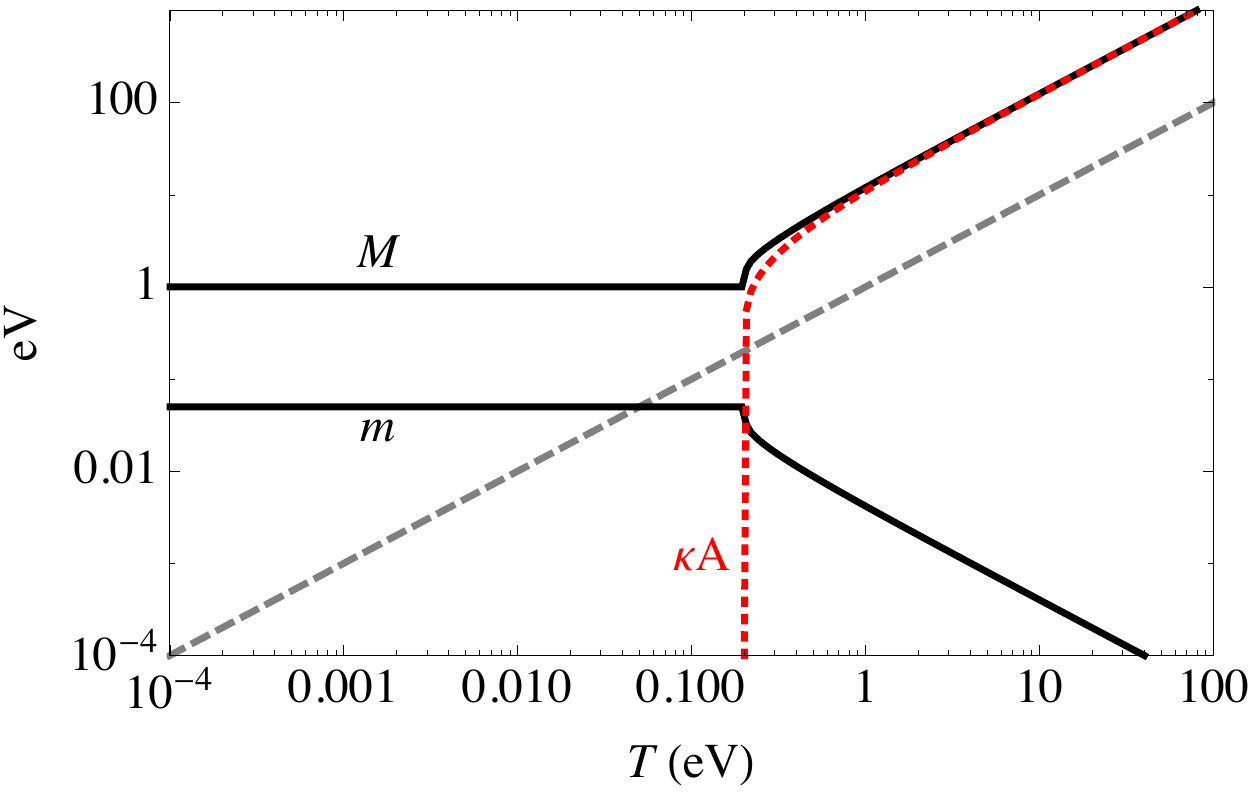}
\caption{Mostly active and mostly sterile neutrino masses, $m$ and $M$ respectively, (solid, black) as functions of the temperature for a logarithmic scalar potential of Eq.~(\ref{eq:Vlog}) with $\Lambda=3.4\times 10^{-2}~\rm eV$. The Majorana mass depends on the scalar field $A$ as in Eq.~(\ref{eq:Mmajorana}) with $m_0=1~\rm eV$ and $\kappa=10^{-6}$. The Dirac mass is taken to be $m_D=0.22~\rm eV$. Also shown is the value of $\kappa A$, (dotted, red). For convenience, the gray, dashed line shows where the mass is equal to the temperature.}\label{fig:m_vs_T}
\end{figure}

We illustrate this behavior in Fig.~\ref{fig:m_vs_T}, showing the neutrino masses and $A$ as functions of temperature for $\Lambda=3.4\times 10^{-2}~\rm eV$, $m_D=0.22~\rm eV$, $m_0=1~\rm eV$, and $\kappa=10^{-6}$.
Although the heavy neutrino's mass chosen to be $1~\rm eV$ today, it is always large compared to the temperature so that its number density is exponentially suppressed and it has no cosmological impact. Correspondingly, the light neutrino mass grows until it reaches a present-day value of $0.05~\rm eV$ around $T=0.1~\rm eV$. Since the active neutrino becomes non-relativistic after its mass becomes independent of temperature it will act as having a mass $m\simeq m_D^2/m_0 = 0.05~\rm eV$ with regards to its impact on cosmology. The active-sterile mixing angle is $\theta\simeq\sqrt{m/M}=0.2$ today and decreases like $T^{-1}$ for $T>0.1~\rm eV$.

\subsection{Including three active neutrinos}
\label{sec:3nus}
Expanding this simple scenario to incorporate three active flavors so that the broad range of neutrino oscillation data can be described is straightforward. The Dirac and Majorana masses become matrices,
\begin{align}
{\cal L}_{\rm mass}&=-{m_D}_{\alpha i}\nu_\alpha N_i-{m_N}_{ij} N_i N_j+{\rm h.c.},
\label{eq:L3flavor}
\end{align}
where $\alpha=e,\mu,\tau$ labels the active flavors while $i,j$ label the sterile neutrinos (there must be at least two to generate the solar and atmospheric mass splittings). For definiteness, we use three sterile neutrinos. In the basis where ${m_N}_{ij}={m_N}_i\delta_{ij}$ is diagonal and taking a Dirac mass matrix of the form
\begin{align}
m_D&=\left(
  \begin{array}{ccc}
    -\sqrt{\frac23}\,\bar m_1 & \sqrt\frac13\,\bar m_2 & 0 \\ 
    \sqrt\frac16\,\bar m_1 & \sqrt\frac13\,\bar m_2 & \sqrt\frac12\,\bar m_3 \\ 
    \sqrt\frac16\,\bar m_1 & \sqrt\frac13\,\bar m_2 & -\sqrt\frac12\,\bar m_3 \\ 
  \end{array}
\right)
\label{eq:m_D}
\end{align}
leads to light neutrinos, $\hat\nu_i$, with masses $m_i=\bar m_i^2/{m_N}_i$ and a light neutrino mixing matrix, $U$, that is approximately tribimaximal.\footnote{Deviating from exact tribimaximal mixing to accommodate the data, in particular $U_{e3}\neq0$, is easy to accomplish by modifying the texture of the Dirac mass matrix slightly.} The effective potential in Eq.~(\ref{eq:Vfull1}) now reads
\begin{align}
V\left(A,T\right)&=\Lambda^4\log\left(1+\left|\frac{A}{\sigma}\right|\right)+\sum_i\frac{m_i^2\left(A\right)T^2}{24},
\label{eq:Vfull3nus}
\end{align}
where the sum runs over each of the light neutrinos.\footnote{For now, the temperature-dependent contributions to the effective potential only matter when each of the light neutrinos is relativistic so we do not have to worry about whether the neutrino background is unstable when they go nonrelativistic, as discussed in~\cite{Afshordi:2005ym}. We return to this point in~\S~\ref{sec:darkenergy}.} Allowing the sterile neutrino masses to depend on the scalar via ${m_N}_i={m_0}_i+\kappa_i A$ results in heavy, mostly sterile neutrinos with masses $M_i\propto T$ at large $T$ and $M_i\simeq {m_0}_i$ at low temperatures. This pattern of couplings results in three active-sterile mixing angles $\theta_i\simeq\sqrt{m_i/M_i}$. Just as in the simple one-flavor case described above, in the early Universe the light, mostly active neutrinos have masses that scale like $T^{-1}$ while the heavy mostly sterile neutrinos' masses go like the temperature, kinematically blocking their production, rendering them cosmologically unimportant.

The mass splittings $\Delta m_{ij}^2=m_i^2-m_j^2$ can be fixed to have present-day values of $\Delta m_{21}^2=\Delta m_{\odot}^2\simeq7.5\times10^{-5}~\rm eV^2$ and $\left|\Delta m_{31}^2\right|=\Delta m_{\rm atm}^2\simeq2.4\times10^{-3}~\rm eV^2$. If we wish to explain the short-baseline reactor anomaly, then we are forced to take $m_1\simeq m_2\sim {\cal O}\left(0.01-0.1~\rm eV\right)$ (and  most naturally $M_1\simeq M_2\sim1~\rm eV$) since the anomalies involve either electron neutrino appearance or disappearance and only $\hat\nu_1$ and $\hat\nu_2$ have appreciable admixtures of $\nu_e$. The third light neutrino, $\hat\nu_3$, can either be taken to be much lighter than the other two, which requires $m_1\simeq m_2\simeq\sqrt{\Delta m_{\rm atm}^2}\simeq0.05~\rm eV$, or roughly degenerate with $\hat\nu_{1,2}$, with $m_{1,2,3}\sim 0.1~\rm eV$. The mixing angle controlling electron neutrino disappearance is then $\theta_{ee}\simeq\theta_{1,2}\sim 0.1$, as needed to explain the reactor anomaly. Electron (anti)neutrino appearance data is more difficult to fit in this model because the $\nu_\mu\to\nu_e$ probability is suppressed by the typical factor of $\theta_{1,2}^4$~\cite{Kopp:2013vaa,*Gonzalez-Garcia:2015qrr} as well as by a further factor of $\theta_{13}^2\ll1$.

\subsection{Acceleron as dark energy}
\label{sec:darkenergy}
In the simple model above, the scalar field $A$ is no longer held away from its true vacuum value. Consequently, it has no connection to dark energy. Here, we describe a way of modifying the model above to allow for one of the light neutrinos to continue holding $A$, the ``acceleron,'' away from its true vacuum value, allowing for an explanation of the dark energy we observe. Explaining dark energy was the original motivation for the mass-varying neutrino scenarios we have been considering~\cite{Fardon:2003eh,Fardon:2005wc}. This dark energy explanation assumes that the true cosmological constant is zero for some reason that does not show up in new particle physics, as in Ref.~\cite{Coleman:1988tj}. We do not address why the true cosmological constant is zero---for a discussion of this problem see   review articles such as Refs.~\cite{Carroll:1991mt,*Carroll:2000fy,*Padmanabhan:2002ji}.

We will begin with the three-flavor model with a logarithmic scalar potential described above, but assume that the $A$-independent contributions to the sterile neutrino Majorana masses are negligible, ${m_0}_i=0$ so that ${m_N}_i=\kappa_i A$ (as before, we work in a basis where the Majorana mass matrix is diagonal). We use a Dirac mass matrix as given in Eq.~(\ref{eq:m_D}) and will be assuming that $\hat\nu_1\propto 2\nu_e-\nu_\mu-\nu_\tau$ and $\hat\nu_2\propto \nu_e+\nu_\mu+\nu_\tau$ are nearly degenerate (as before we assume a tribimaximal mixing matrix). We now add Majorana masses for the active neutrinos of the form
\begin{align}
{\cal L}_{\rm mass}\supset-\mu\left[\nu_e\nu_e+\frac12\left(\nu_\mu+\nu_\tau\right)^2\right]+{\rm h.c.}
\end{align}
This gives a common Majorana mass of $\mu$ for $\hat\nu_1$ and $\hat\nu_2$. Such masses could be generated by integrating out another set of Majorana sterile fermions (that do not couple strongly to $A$). Given a hierarchy between the Dirac masses and the sterile Majorana masses, the heavy neutrino masses are simply $M_i\simeq\kappa_i A$ and the masses of the the light, mostly active neutrinos are $m_1\simeq m_2\simeq\left|\mu-\bar m_{1,2}^2/\kappa_{1,2}A\right|$ and $m_3\simeq\bar m_3^2/\kappa_3A$. The active-sterile mixing angles are $\theta_i\simeq \bar m_i/M_i$.

An important consideration in in this scenario is the instability to collapse which occurs when a neutrino becomes nonrelativistic~\cite{Stephenson:1996qj,Afshordi:2005ym}. These nonrelativistic neutrinos form ``nuggets'' and are no longer a cosmologically relevant background when computing the effective potential of the acceleron. Hence, they no longer help to ``hold up'' the acceleron vacuum expectation value (vev). In the Appendix we show details of the calculation of the temperature at which this instability develops.

\begin{figure}
\includegraphics[width=\linewidth]{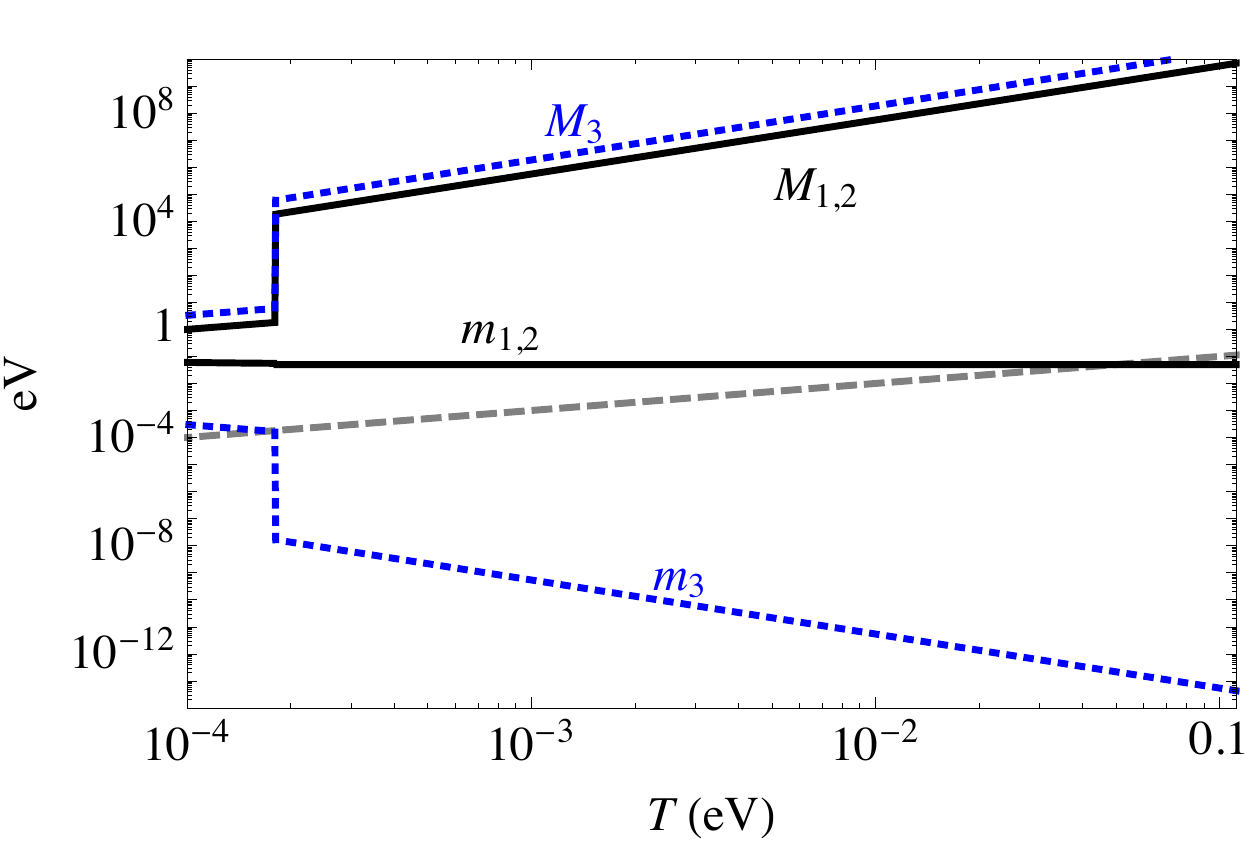}
\caption{Mostly active and mostly sterile neutrino masses, $m_i$ and $M_i$ respectively, as functions of the temperature for a logarithmic scalar potential of Eq.~(\ref{eq:Vlog}), with $\Lambda=9.3\times 10^{-5}~\rm eV$, in the case where $\hat\nu_{1,2}$ have (equal) nonzero, $A$-independent Majorana masses, $\mu=-0.05~\rm eV$. The sterile neutrino Majorana masses are ${m_N}_i=\kappa_i A$ with $\kappa_{1,2}=1$, $\kappa_3=3.3$. The Dirac masses are $\bar m_{1,2}=0.1~\rm eV$, $\bar m_3=0.031~\rm eV$. The gray, dashed line shows where the mass is equal to the temperature. $\hat\nu_{1,2}$ no longer contribute to the effective $A$ potential when $T\lesssim 1.8\times10^{-4}~\rm eV$ (see Appendix for details).}\label{fig:m_vs_T2}
\end{figure}

Because of this, we work with an inverted hierarchy $m_3\ll m_{1,2}$; the lightest neutrino is responsible for supporting $A$ today, while the heavier two, which have an appreciable admixture of electron neutrino, provide for oscillations into sterile neutrinos with $\Delta m^2\simeq 1~\rm eV^2$. In this case, at high temperatures when all of the light neutrinos are relativistic, the sum in the effective potential of Eq.~(\ref{eq:Vfull3nus}) runs over $i=1,2,3$ and $\hat\nu_1$ and $\hat\nu_2$ are dominantly responsible for keeping the acceleron away from its true minimum. Minimizing the potential then leads to $A\propto T^2$. At some point after $\hat\nu_1$ and $\hat\nu_2$ go nonrelativistic, the sum in Eq.~(\ref{eq:Vfull3nus}) only includes $i=3$. The value of $A$ that minimizes the effective potential is determined by $\hat\nu_3$, so that $A\propto T$. We show the masses of the neutrinos as functions of the temperature in Fig.~\ref{fig:m_vs_T2}, setting $\mu=-0.05~\rm eV$, $\bar m_{1,2}=0.1~\rm eV$, $\bar m_3=0.031~\rm eV$, $\kappa_{1,2}=1$, $\kappa_3=3.3$, and $\Lambda=9.3\times 10^{-5}~\rm eV$. As shown in the Appendix, given these parameters, $\hat\nu_{1,2}$ stop supporting the $A$ vev at $T\simeq 1.8\times10^{-4}~\rm eV$. Note that due to the presence of a Majorana mass for the active neutrinos, the conclusion of Ref.~\cite{Spitzer:2006hm} where a cascade of ``nugget'' formation occurs after the heaviest neutrino goes nonrelativistic does not apply.

To get a vacuum energy density $\sim10^{-11}~\rm eV^4$ requires a very large value for the logarithm in the scalar potential. This could perhaps be most natural in a scenario where a dilaton-like field, $\phi$, is the dynamical origin of the $A$-dependent contribution to the Majorana mass and $A=A_0\exp\left(\phi^2/f^2\right)$. Another possibility is a more conventional quintessence model with an acceleron Compton wavelength on the order of the Hubble scale.  In such a model neutrino masses evolve but do not collapse to form nuggets, even when nonrelativistic. In this case the light neutrinos can be quite massive today, of order an eV.\footnote{The neutrinos masses locally today in this case would be smaller due to the relative overdensity of neutrinos in our galaxy cluster.} The effects of such massive MaVaNs on CMB and structure has been explored in Ref.~\cite{Ghalsasi:2014mja}.

\section{A logarithmic potential and active neutrinos at an $\rm eV$}
\label{sec:mass}

We now turn to the question of whether the observed {\it active} neutrinos can have masses around an $\rm eV$ today.
	
	Neutrinos can free stream over cosmological distances and thus damp perturbations on a scales smaller than the free streaming scale. For standard neutrinos with the sum of their masses $\Sigma m_{\nu}\sim1~\rm eV$ this effect is observable via suppression of the matter power spectrum on scales smaller than free streaming scale. The larger the neutrino masses the larger the energy density contained in them around matter radiation equality when structures start to grow. Hence, the damping of the matter power spectrum on scales smaller than the free streaming scale is more pronounced as the masses of the neutrinos is increased.  Large scale structure surveys which are sensitive the matter power spectrum can therefore put an upper limit on $\Sigma m_{\nu}$ (for details, see, e.g. Ref.~\cite{Lesgourgues:2012uu}).
	
	However, mass-varying neutrinos can act like massless neutrinos during and after matter-radiation equality and become massive at much lower redshifts. Here we construct a phenomenologically viable scenario using the framework described in~\S~\ref{sec:logpotential}, with a logarithmic scalar potential for the field $A$. For simplicity let us consider the case of one active and one sterile neutrino, leading to an effective potential like that in Eqs.~(\ref{eq:Vfull1}) and (\ref{eq:Vfull2}). Generalizing this, as in~\S~\ref{sec:3nus}, to three flavors (that are nearly degenerate since we will be describing active neutrinos with a mass around an eV today) is straightforward. We assume that $m_{0}=0$ in Eq.~(\ref{eq:Mmajorana}). At high temperatures when the light neutrino is relativistic, minimizing the potential in Eq.~(\ref{eq:Vfull2}) leads to a light, mostly active, neutrino of mass $m\left(T\right)\simeq\sqrt{12}\Lambda^{2}/T$ and a heavy, mostly sterile neutrino with mass $M\left(T\right)\simeq m_D^2T/\sqrt{12}\Lambda^{2}$. As mentioned in the previous section the light neutrinos condense to form ``nuggets'' after going nonrelativistic and stop supporting the scalar field, which settles to its minimum at $A=0$. We show in the Appendix that this occurs when $T\simeq m\left(T\right)/10$. The energy density stored in the cosmological neutrinos is now stored in the nuggets which redshift like like matter.  The active and sterile neutrinos now form a Dirac fermion of mass $m_{D}$ which is independent of the temperature and can be much larger than the mass of the light neutrinos when they went nonrelativistic.

\begin{figure}
\includegraphics[width=\linewidth]{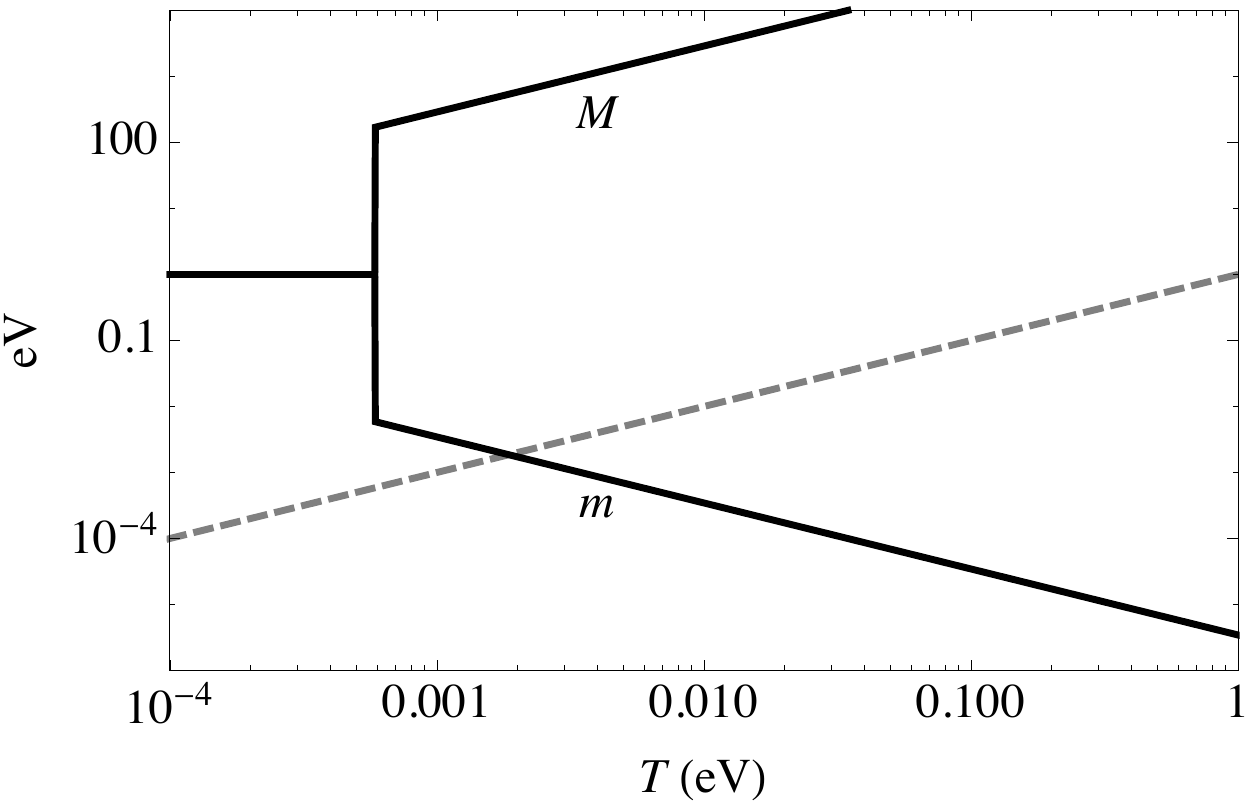}
\caption{The masses of the light, mostly active and heavy, mostly sterile neutrinos, $m$ and $M$ respectively, in a single active flavor scheme as a function of temperature for $m_0=0$, $m_D=1~\rm eV$, and $\Lambda = 10^{-3}~\rm eV$. When $T\simeq m/10$, the light neutrino ceases supporting the $A$ vev and the sterile Majorana mass vanishes. The sterile and active neutrinos then form a Dirac fermion.}\label{fig:m_vs_T3}
\end{figure}

\begin{figure}
\includegraphics[width=\linewidth]{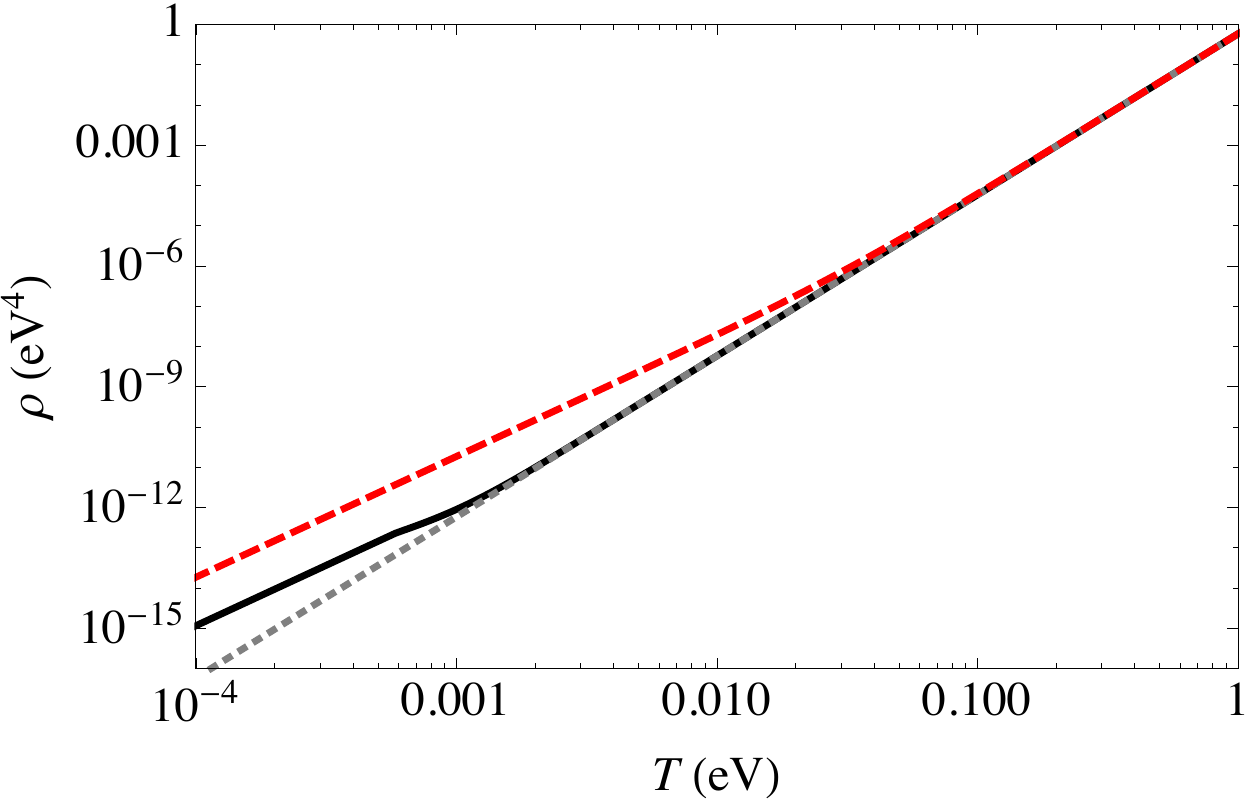}
\caption{The neutrino energy density as a function of temperature for $m_0=0$, $m_D=1~\rm eV$, and $\Lambda = 10^{-3}~\rm eV$ (solid, black). Also show are the energy densities for {\it constant} mass neutrinos of mass $0.1~\rm eV$ (dashed, red) and zero (dotted, gray).}\label{fig:rho_vs_T}
\end{figure}

	In order to illustrate our point let us consider $m_{D} = 1~\rm eV$ and $\Lambda = 10^{-3}~\rm eV$. We plot the light neutrino mass vs. temperature in Fig.~\ref{fig:m_vs_T3}. As can be seen, the neutrino acts effectively massless around matter-radiation equality. After the neutrino becomes nonrelativistic at $T\sim\Lambda$ it stops supporting the acceleron. The sterile and active neutrinos now form a Dirac fermion of mass $m_{D}$ which is now independent of temperature. This is the mass that is relevant for terrestrial neutrino mass experiments, e.g., searches for endpoints in $\beta$-decay spectra. However the relevant mass of the neutrino from the standpoint of experiments which measure the matter power spectrum can be approximated to be the mass when the neutrino becomes nonrelativistic, i.e. $m\sim \Lambda\ll m_D$. 
	
To see this in further detail we show the energy density stored in the neutrinos (and ``nuggets'' after their formation) as the temperature varies in Fig.~\ref{fig:rho_vs_T}. We also show the energy density of a {\it constant} $0.1~\rm eV$ mass neutrino, which is roughly the upper limit on $\sum m_\nu/3$ from cosmological observations. We see that the energy density in the mass-varying neutrino in this case is never greater than that in the constant $0.3~\rm eV$ neutrino. Indeed, the {\it temperature-dependent} energy density in the mass varying neutrino is roughly what one would find for a constant mass neutrino of mass $\sim\Lambda=10^{-3}~\rm eV$.

A signal of this mechanism would be detection of an eV scale neutrino mass in tritium beta decay of \cite{Robertson:2013ziv,Doe:2013jfe}, but with no signature in double beta decay or in searches for massive cosmological neutrinos \cite{Weinberg:1962zza,Betts:2013uya,Long:2014zva}, since the cosmological neutrinos would be clustered and it is unlikely that we would be inside a nugget.

\section{A supersymmetric potential}
\label{sec:susy}
MaVaNs theories can be made supersymmetric~\cite{Fardon:2005wc,Takahashi:2005kw}, which is well motivated by the necessity of including a light scalar. In this case, however, the scalar potential is constrained to be quadratic in the acceleron field, and not logarithmic as we have so far considered. Dark energy may be obtained via the acceleron-sneutrino coupling, which   creates an effective sneutrino potential with a minimum which differs from the vacuum configuration.

Let us begin by briefly describing the supersymmetric MaVaNs theory of a single active neutrino. The superpotential of this theory after electroweak symmetry breaking is
\begin{align}
W=\kappa\, ann+{m_D}\nu n,
\end{align}
where $\nu$, $n$, and $a$ are the superfields containing the active neutrino, sterile neutrino, and acceleron respectively. Including both SUSY-preserving and -breaking interactions, the scalar potential is
\begin{equation}
\begin{aligned}
V_{\rm scalar}=&{\mu_{\tilde N}}^2\big|\tilde N\big|^2+m_A^2\big|A\big|^2+4\kappa^2\big|A\big|^2\big|\tilde N\big|^2
\\
&\quad+\kappa^2\big|\tilde N\big|^4+{\rm const.},
\label{eq:Vsinglscalar}
\end{aligned}
\end{equation}
with $\tilde N$ the sterile sneutrino and the constant such that the true minimum of the potential is at $V_{\rm scalar}=0$. Radiative corrections drive ${\mu_{\tilde N}}^2$ to negative values of order $-m_D^2$, which means that $\langle\tilde N\rangle=\sqrt{-{\mu_{\tilde N}}^2/2\kappa^2}\sim {\cal O}\left(m_D/\kappa\right)$ at the true minimum of the potential. As before, the acceleron field is driven to large values at finite temperature due to the contribution to the effective potential from the light neutrino density. If $A>\sqrt{-{\mu_{\tilde N}}^2/4\kappa^2}\sim{\cal O}\left(m_D/\kappa\right)$ then the sneutrino is trapped at a local minimum with energy density ${\mu_{\tilde N}}^4/4\kappa^2\sim{\cal O}\left(m_D^4/\kappa^2\right)$.

The effective potential that determines the finite-temperature value of the acceleron is
\begin{equation}
\begin{aligned}
V\left(A,T\right)&=m_A^2\big|A\big|^2+\frac{m^2\left(A\right)T^2}{24}
\\
&=m_A^2\big|A\big|^2+\frac{m_D^4T^2}{24\kappa^2\big|A\big|^2},
\end{aligned}
\end{equation}
where we have assumed that the neutrino mass matrix is of the seesaw form. Minimizing this results in
\begin{align}
A\left(T\right)&=\frac{m_D}{24^{1/4}}\sqrt{\frac{T}{\kappa m_A}},
\label{eq:Aquadpotential}
\end{align}
in contrast to the case of a logarithmic potential where $A\propto T$.

The acceleron mass term receives SUSY-breaking radiative corrections and, in the absence of fine tuning, we expect $m_A^2\gtrsim\kappa^2 m_D^2$. Therefore, to keep the sneutrino in the false minimum today when $T=T_0\simeq 10^{-4}~\rm eV$ requires that $m_D\lesssim T_0$ in a natural theory.

We now discuss how to extend this treatment to the case of three active neutrinos which means that the couplings $\kappa$ and $m_D$ are now matrices. As in Ref.~\cite{Fardon:2005wc}, a viable set of parameters involves the acceleron vev being held up by the lightest neutrino, which remains relativistic today. To connect with short-baseline anomalies, it is phenomenologically motivated to take an inverted mass hierarchy, so that the smallest neutrino mass is $m_3$. Like in~\S~\ref{sec:3nus}, we work in a basis where the matrix $\kappa_{ij}=\kappa_i\delta_{ij}$ is diagonal and the Dirac mass matrix is of the form in Eq.~(\ref{eq:m_D}). We take the Dirac mass corresponding to the lightest neutrino to be $\bar m_3=10^{-5}~\rm eV$ and $\kappa_3=10^{-5}$. This gives a present-day dark energy density of the correct order of magnitude $\sim 10^{-11}~\rm eV^4$. 

To avoid having to fine-tune away large radiative contributions to the acceleron mass coming from the neutrinos of larger mass, we take $\kappa_{1,2}\ll\kappa_3$. (We will state more precise values for $\kappa_{1,2}$ below.) If this is the end of the story for the neutrino masses, then $\hat\nu_{1,2}$ are essentially Dirac fermions containing the active and sterile neutrinos orthogonal to $\hat\nu_3$ and $\hat N_3$, which does not allow for any mass splitting in this system $\sim 1~\rm eV$. To fix this we can add an acceleron-independent contribution to the sterile neutrino masses as we did in~\S~\ref{sec:logpotential} through the superpotential term $W\supset {m_N}_{ij}n_in_j$ which leads to terms in the scalar potential ${m_N}_{ij}^2\tilde N_i\tilde N_j^\ast$. If we assume that the terms in this mass matrix involving $i,j=3$ are suppressed and the nonzero eigenvalues of this matrix are ${\cal O}\left(1~\rm eV\right)$, then there will be a pair of a sterile neutrinos mass of around an eV, allowing for a oscillations of the active neutrinos with $\Delta m^2\simeq 1~\rm eV^2$. Furthermore, the contribution to the scalar potential from ${m_N}_{ij}$ pushes $\tilde N_{1,2}\to 0$, simplifying the analysis of the scalar potential which is then essentially that of a single sterile state as in Eq.~(\ref{eq:Vsinglscalar}), involving only the acceleron and $\tilde N_3$. Describing atmospheric neutrino oscillations then requires $\bar m_1\simeq\bar m_2\simeq 0.22~\rm eV$ so that $m_1\simeq m_2\simeq 0.05~\rm eV$.

Because, as we see in Eq.~(\ref{eq:Aquadpotential}), $A\propto\sqrt T$, the sterile neutrinos are not kinematically forbidden from being produced in the early Universe, unlike the case of the logarithmic potential. This could be in strong conflict with cosmological bounds and furthermore a large sterile neutrino density will affect the acceleron potential, driving the acceleron to small values, ruining this mechanism as an explanation of the dark energy. This can be simply avoided by a Planck-suppressed coupling of the acceleron to electrons, ${\cal L}\supset \beta_e(m_e/M_{\rm Pl})A\bar ee$, where $\beta_e$ is a coupling and $M_{\rm Pl}$ is the Planck mass, as described in Ref.~\cite{Weiner:2005ac}. Fifth force tests limit $\left|\beta_e\right|\lesssim 4$. At temperatures above the electron mass, the (mostly) neutrino masses are roughly
\begin{align}
M_i\simeq3~{\rm MeV} \beta_e\left(\frac{\kappa_i}{10^{-10}}\right)\left(\frac{T}{1~\rm MeV}\right)^2\left(\frac{10^{-11}~\rm eV}{m_A}\right)^2.
\end{align}
As previously mentioned, the acceleron receives quantum corrections and in a natural theory we expect $m_A^2\gtrsim\sum_i\kappa_i^2 \bar m_i^2$. For the values of $\bar m_3$ and $\kappa_3$ given above, an acceleron mass of around $10^{-11}~\rm eV$ requires a modest 10\% fine tuning and limits $\kappa_{1,2}\lesssim{\cal O}\left(10^{-4}\kappa_3\right)\sim 10^{-9}$. Production of sterile neutrinos at $T\gtrsim \rm MeV$ is therefore suppressed if $\beta_e\sim{\cal O}\left(1\right)$, which is enough to bring the scenario into agreement with cosmological limits. For further details and constraints, see Ref.~\cite{Weiner:2005ac}.

\section{Conclusions}
\label{sec:conclusions}
In this paper we have described several situations involving mass-varying neutrinos that allow for either active neutrinos or sterile neutrinos with a large active-sterile mixing to have masses around an eV and yet still be compatible with strong limits from cosmological observations. Along with ``secret'' neutrino interactions~\cite{Hannestad:2013ana,*Dasgupta:2013zpn,*Bringmann:2013vra,*Saviano:2014esa,*Chu:2015ipa,Archidiacono:2014nda,Archidiacono:2016kkh}, this possibility illustrates the necessity of combining cosmological probes of neutrino properties with terrestrial experiments. Combining both probes allows us to {\it test} whether neutrinos have richer structure than expected in ways that cosmological observations or terrestrial experiments alone cannot.

MaVaNs are motivated by attempts to understand dark energy. It is interesting that they can also modify neutrino cosmology to allow recent hints for eV-scale sterile neutrinos to be reconciled with cosmological observations, or for the active neutrinos to have masses within the reach of near future experiments today. In addition, it is worth mentioning that unlike the case of ``secret'' neutrino interactions, the (mostly) sterile neutrinos in this scenario with a logarithmic potential are kinematically forbidden from being produced at late times, and therefore do not suffer from the problem that they are produced in late-time collisions, upsetting agreement with cosmological observations~\cite{Mirizzi:2014ama}. 

The requirement of a light scalar field for the MaVaNs scenario suggests that this sector could be supersymmetric. Properly supersymmetrizing the theory  adds additional constraints and requires introducing very weak couplings of this light field to charged fermions in order to make the sterile neutrinos heavy at early times. Besides cosmological observations and neutrino experiments, these weak couplings offer perhaps the best way of testing the scenario, through, e.g., searches for fifth forces at large distance scales~\cite{Adelberger:2003zx} or electron-density--dependent neutrino masses~\cite{Kaplan:2004dq,Zurek:2004vd}.

Interactions between light scalars and active neutrinos can prevent active neutrinos from freely streaming and lead to observable signatures in the CMB~\cite{Friedland:2007vv,*Basboll:2008fx,*Smith:2011es,*Cyr-Racine:2013jua,*Archidiacono:2013dua}. In the scenarios we have considered, the active neutrino-scalar coupling is too small to be constrained by these considerations. It would be interesting to study such CMB signatures and their complementarity with other terrestrial observables in nonstandard neutrino scenarios.

We have extended earlier work on MaVaNs to take into account the effects of neutrino clustering, to add additional contributions to neutrino mass and additional couplings  in the scalar sector. We have been able to exhibit models in which eV mass sterile neutrinos appear in present day neutrino oscillation experiments but do not affect precision cosmology, and in which active neutrinos could have eV scale masses today but not in the early universe.

Neutrino masses, inflation, dark matter, baryogenesis, and dark energy all show that there must be new physics beyond the standard model. We do not know the energy scale, but neutrino masses and dark energy   indicate a physical scale of order $10^{-4}-10^{-2}$ eV. It would be remiss of us to simply adopt theoretical prejudice  and assume that such new physics  does not involve any new light particles. If there is such new physics, then  precision cosmology and laboratory measurements are not necessarily simply different ways of measuring the same neutrino properties. As an illustration of the possibilities, in this paper we have explored models containing light sterile neutrinos coupled to  a  light scalar in which laboratory and cosmological measurements which would give seemingly inconsistent results can instead be interpreted as evidence for a new light sector, which could be the origin of dark energy.

\appendix\label{appendix}
\section{Instability in MaVaNs}
	
As pointed out in Ref.~\cite{Afshordi:2005ym}, there is a inherent instability in the neutrino-acceleron fluid when the neutrinos go nonrelativistic. The end result of the instability is that the neutrinos condense to form ``neutrino nuggets'' which then redshift like matter. To see this easily, one can treat the neutrino-acceleron fluid hydrodynamically. 

For now, assume that there is just one light neutrino and that its mass is inversely proportional to the scalar field, $m\propto 1/A$. When the neutrinos are nonrelativistic, the effective potential from Eqs.~(\ref{eq:deltaV}) and (\ref{eq:Vfull}) is
\begin{align}
V\left(A,T\right)&=\Lambda^4\log\left|\frac{\sigma^\prime}{m}\right|+m\, n_\nu,
\label{eq:Vfullnr}
\end{align}
where $\sigma^\prime$ is a scale and $n_\nu$ is the light neutrino number density. Minimizing this potential with respect to $A$ gives
\begin{align}
n_{\nu} = -\frac{\partial V_{0}}{\partial m} \simeq \frac{\Lambda^{4}}{m},
\end{align}
so that the light neutrino energy density is roughly constant, $\rho_\nu\simeq m\, n_\nu\simeq\Lambda^{4}$.

Now we calculate the speed of sound of the neutrino-acceleron fluid. The fluid's pressure is $P = -\rho_{a} + w_{\nu}\rho_{\nu}$ where $\rho_{a} = V_{0}$ is the energy density stored in the acceleron field and $w_\nu$ is the neutrino's equation of state parameter. Since the neutrino is nonrelativistic, $w_{\nu}\simeq 0$. The energy density of the fluid is given by $\rho = \rho_{\nu}+\rho_{a}$. The speed of sound is related to the rate of change of the pressure and energy density,
\begin{align}
c^{2}_{s} &= \frac{\dot{P}}{\dot{\rho}} \simeq -\frac{\dot{\rho_{a}}}{\dot{\rho_{a}}} = -1,
\end{align}
where we have used the fact that $\rho_\nu$ is approximately constant. Because density perturbations evolve in time as $e^{-ic_s t}$, an imaginary component to $c_s$ signals an instability, and we see that the neutrino-acceleron fluid is unstable when the neutrinos are nonrelativistic.

A more sophisticated picture can be obtained from a kinetic theory treatment where the speed of sound is determined by~\cite{Afshordi:2005ym}
\begin{equation}
\begin{aligned}
c^{2}_{s} &= \frac{1}{m}\frac{\partial V_{0}}{\partial m}\left(\frac{\partial ^{2} V_{0}}{\partial m^{2}}\right)^{-1}\times \\ &\left[1+12.9\left(c^{2}_{s}+c^{-2}_{s}-2\right)\left(\frac{T}{m}\right)^{2}+ O\left(\frac{T}{m}\right)^{4}\right].
\label{eq:speed}
\end{aligned}
\end{equation}
Again, the fluid becomes unstable when the speed of sound, as found by this relationship, develops an imaginary component. Below, we use this expression to determine the temperature at which the neutrino-acceleron fluid becomes unstable for the scenarios considered in~\S\S~\ref{sec:darkenergy} and \ref{sec:mass}.

In~\S~\ref{sec:darkenergy} the heavy active neutrinos masses get contributions from the active Majorana mass $\mu$ which are independent of acceleron vev as well as contributions that depend on the acceleron vev. Taking $m_1\simeq m_2\equiv m$, we rewrite the logarithmic part of the acceleron potential in terms of the neutrino mass as
\begin{align}
V\left(A,T\right)&=V_0+m\, n_\nu=\Lambda^4\log\left|\frac{\mu_{0}}{m+\mu}\right|+m\, n_\nu.
\label{eq:Vfullnr}
\end{align}
Here, $n_\nu$ is the sum of the number densities of the two nearly degenerate neutrinos $\hat\nu_{1,2}$. This gives
\begin{align}
\frac{1}{m}\frac{\partial V_{0}}{\partial m}\left(\frac{\partial ^{2} V_{0}}{\partial m^{2}}\right)^{-1}&=-\frac{1}{m}\frac{\bar m_{1,2}^2}{\kappa_{1,2} A}.
\end{align}
Minimizing the effective potential implies that
\begin{align}
\frac{\bar m_{1,2}^2}{\kappa_{1,2} A}\simeq\frac{\pi^2}{3\zeta\left(3\right)}\frac{\Lambda^4}{T^3}.
\end{align}
$\Lambda$ is determined by the present-day mass of the lightest neutrino and temperature, $\Lambda^4=m_3^2\left(T_0\right)T_0^2/12$ and we can approximate $m$ with $\left|\mu\right|$. Thus, we can write Eq.~(\ref{eq:speed}) as
\begin{equation}
\begin{aligned}
c^{2}_{s} &=-\frac{\pi^2}{36\zeta\left(3\right)}\frac{m_3^2\left(T_0\right)T_0^2}{\left|\mu\right|T^3}\times \\ &\left[1+12.9\left(c^{2}_{s}+c^{-2}_{s}-2\right)\left(\frac{T}{m}\right)^{2}+ O\left(\frac{T}{m}\right)^{4}\right]
\\
&=-0.23\times\frac{m_3^2\left(T_0\right)T_0^2}{\left|\mu\right|^4}\left(\frac{m}{T}\right)^{3}\times \\ &\left[1+12.9\left(c^{2}_{s}+c^{-2}_{s}-2\right)\left(\frac{T}{m}\right)^{2}+ O\left(\frac{T}{m}\right)^{4}\right].
\end{aligned}
\end{equation}
For the parameter values specified in~\S~\ref{sec:darkenergy} we get the temperature at which the nuggets are formed to be $T\simeq 1.8\times 10^{-4} ~\rm eV$.

A similar calculation can be done for the scenario described in~\S~\ref{sec:mass}. In that case there is no active Majorana mass and
\begin{align}
\frac{1}{m}\frac{\partial V_{0}}{\partial m}\left(\frac{\partial ^{2} V_{0}}{\partial m^{2}}\right)^{-1}&=-1.
\end{align}
As before, Eq.~(\ref{eq:speed}) can then be solved to determine when nuggets form. We find that this occurs at the temperature $T\simeq m/10$.

\bibliography{mavans}

\end{document}